\documentclass[%
 reprint,prl,
superscriptaddress,
showpacs,
 amsmath,amssymb,
 aps,
]{revtex4-1}

\usepackage[caption=false]{subfig}
\usepackage{graphicx}
\usepackage{dcolumn}
\usepackage{bm}
\usepackage{hyperref}

\bibstyle{aipauth4-1}

\newcommand{\etal}{\textit{et al.}}

\begin{document}

\title{The Elasticity of Nuclear Pasta}

\author{M. E. Caplan}
 \email{matthew.caplan@mcgill.ca}
 \affiliation{%
 McGill Space Institute, McGill University, 3600 Rue University, Montreal, QC Canada H3A 2T8 }
\author{A. S. Schneider}%
\affiliation{TAPIR, Walter Burke Institute for Theoretical Physics, California Institute of Technology, Pasadena, CA 91125, USA}
\author{C. J. Horowitz}
\email{horowit@indiana.edu}
\affiliation{Nuclear Theory Center, Indiana University, Bloomington, IN 47401, USA}
\date{\today}

\begin{abstract}

The elastic properties of neutron star crusts are relevant for a variety of currently observable or near-future electromagnetic and gravitational wave phenomena.  
These phenomena may depend on the elastic properties of nuclear pasta found in the inner crust. 
We present large scale classical molecular dynamics simulations where we deform nuclear pasta. We simulate idealized samples of nuclear pasta and describe their breaking mechanism. We also deform nuclear pasta that is arranged into many domains, similar to what is known for the ions in neutron star crusts. Our results show that nuclear pasta may be the strongest known material, perhaps with a shear modulus of $10^{30}\,\text{ergs/cm}^3$ and breaking strain greater than 0.1. 
\end{abstract}

\maketitle

\textit{Introduction.} The breaking strain of materials in neutron star (NS) crusts are relevant for a variety of electromagnetic and gravitational wave phenomena. Crust breaking may occur in magnetar outbursts \cite{0004-637X-561-2-980}, resonant crust shattering in NS mergers \cite{PhysRevLett.108.011102}, and in the starquake model for pulsar glitches \cite{Ruderman,Chugunov2010}. Likewise, the shear modulus of crust matter may affect the  oscillation frequency of magnetar flares \cite{Strohmayer2004}, the damping of r-modes \cite{Haskell2012}, and determine the height and lifetime of mountains on a NS \cite{Chugunov:10,Mahmoodifar:13}.

The crust comprises the outermost kilometer of the NS, where the density goes from near vacuum at the surface to nuclear density ($10^{14}\,\text{g/cm}^3$) at the base. The outer crust is a bcc lattice of nuclei embedded in a gas of degenerate electrons, which becomes increasingly neutron rich with depth \cite{Chamel2008,astromaterials}. At the base of the inner crust the separation between nuclei becomes comparable to nuclei radii and nucleons rearrange themselves into complex shapes known as nuclear pasta \cite{PhysRevLett.50.2066,Hashimoto1984}. The pasta layer may be 100-250\,m thick, but due to its high density may contain half the total mass of the crust or more \cite{Newton2013}. It is therefore essential to understand the elastic properties of this material, as it may play an important role in crust breaking and be the dominant source of continuous gravitational waves that aLIGO is now searching for \cite{PhysRevD.96.122006}.

Past works have studied elastic properties (breaking strain and shear modulus) of the ion crust with both molecular dynamics (MD) and analytic methods \cite{Chamel2008, astromaterials, Caplan2018, PhysRevLett.102.191102, Chugunov2010, Kobyakov2015}.
The ion crust is understood to be a polycrystalline bcc lattice, \textit{i.e.}, composed of microscopic domains with different orientations. Thus, over sufficiently large length scales the crust is treated as an isotropic material whose properties are found by angle averaging bcc lattice properties \cite{Kobyakov2015}. MD simulations suggest that domains lead to an effective angle-averaged shear modulus \cite{Chugunov2010,PhysRevLett.102.191102} and that high pressure prevents voids from forming and fractures from propagating, thus, leading to a relatively high breaking strain $\epsilon \approx 0.1$ \cite{PhysRevLett.102.191102}.

Presently, the pasta is less well understood. Unlike the ion lattice pasta is not a crystal but, rather, a liquid crystal. 
Pethick \& Potekhin, Ref. \cite{PETHICK19987}, were the first to study the elasticity of nuclear pasta. They considered the energy of deformation $E_d$ of parallel pasta plates to be
\begin{equation}
E_d = \frac{B}{2} \left[ \frac{\partial u }{\partial z} - \frac{1}{2}( \nabla_\perp u)^2   \right]^2 + \frac{K_1}{2}  ( \nabla_\perp^2 u)^2\,,
\end{equation}
where $B$ and $K_1$ are elastic constants and $u$ the displacement between plates along the $z$ axis \cite{PETHICK19987}. The first term is the energy due to the separation of the plates. The $(\nabla_\perp u)^2$ term is due to rotational invariance, as out-of-plane shear is rotationally equivalent to changing the plate spacing. The last term is due to splay deformations (plate curvature).

These analytic techniques are difficult to apply to asymmetric pasta or pasta lacking long range order. Even short range disorder such as helicoidal defects (filaments connecting the plates) may support shear stresses between plates. These may be difficult to predict analytically, but can be studied using MD simulations. 
The same is true for the effect of domains and their boundaries in nuclear pasta on crust elasticity.

Watanabe et al. compare the typical thermal energy to the energy of deformation of nuclear pasta and find that pasta deforms on length scales of order ten times its spacing \cite{WATANABE2000455}. This result is supported by quantum mechanical simulations that suggest that the pasta, like the ion crust, does not have a uniform orientation across the star \cite{Newton2018}. Rather, nuclear pasta should be composed of many microscopic domains with distinct orientations. Thus, calculations of the elastic properties of the crust should consider `polycrystalline' nuclear pasta with many domains. Currently, such studies are limited to MD simulations since those can simulate large volumes with multiple domains \cite{Schneider:18} while quantum mechanical simulations are limited a few thousand nucleons \cite{PhysRevC.95.055804}.

In this Letter we discuss MD simulations (1) deforming nuclear pasta plates, to understand the breaking of its elementary units, and (2) deforming a large nuclear pasta system with multiple domains, to understand angle averaged elastic properties of nuclear pasta.

\textit{Simulations.} Our simulations are performed using the Indiana University Molecular Dynamics code v6.3.1 (IUMD), which has been used extensively to simulate nuclear pasta \cite{PhysRevC.69.045804,PhysRevC.88.065807,astromaterials}. In our model, two nucleons $i$ and $j$ separated by a distance $r$ interact via the two-body potential
\begin{equation}
 V_{ij}(r)=ae^{-r^2/\Lambda}+[b \pm c]e^{-r^2/2\Lambda}+\frac{e_i e_j }{r}e^{-r/\lambda}\,.
\end{equation}
The parameters $a$, $b$, $c$, and $\Lambda$ are chosen to reproduce the binding energy of nuclei, pure neutron matter, and symmetric nuclear matter while $\lambda$ is the Thomas-Fermi screening length, fixed at 10 fm for simplicity \cite{PhysRevC.69.045804}. 
The $b+c$ ($b-c$) term sets a weak (strong) attraction between like (alike) nucleons. The electric charges $e_i$ and $e_j$ produce long range Coulomb repulsion between protons. All systems are simulated at temperature $T=1$ MeV and nucleon number density $n=0.05\,\text{fm}^{-3}$.

Our first set of runs contain 102\,400 nucleons at a proton fraction $Y_P=0.4$, where pasta is expected to form `lasagna' plates \cite{PhysRevC.88.065807,astromaterials}. This proton fraction is higher than expected in the inner crust, but is where our model forms pasta; at more realistic proton fractions the classical model produces a gas of protons and neutrons \cite{PhysRevC.90.055805}.

We initialize our simulation with a cubic volume (side length $l_0$) with periodic boundary conditions and apply volume preserving deformations. We apply constant extensional strain along axis $r$ and contract along the other two, $s$ and $t$, by adjusting the box boundary position slightly after each MD timestep, \textit{i.e.},
\begin{equation}
l_r(t) = l_0 ( 1 + \dot{\epsilon} t) \quad\text{and}\quad
l_{s}(t) = l_{t}(t) = \frac{l_0}{\sqrt{1+\dot{\epsilon}t}}.
\end{equation}
where $\dot\epsilon$ is the strain rate and $r$, $s$, and $t$ are permutations of $x$, $y$, and $z$. Nucleons respond dynamically to the changing simulation volume and the induced stresses are calculated using
\begin{equation}\label{strain}
 \sigma_{\alpha\beta} = \frac{1}{V} \sum_{i} \left[-m u_\alpha^{(i)} u_\beta^{(j)} + \sum_{i < j } ( x_\beta^{(i)} - x_\alpha^{(i)}) f_\beta^{(ij)}\right].
\end{equation}
Above, $V$ is the simulation volume, $m$ the nucleon mass, $x_\alpha^{(i)}$ ($u_\alpha^{(i)}$) the $\alpha$ component of the position (velocity) of nucleon $i$ and $f_\beta^{(ij)}$ the $\beta$ component of the force nucleon $i$ exerts on nucleon $j$. Repeated (mixed) indices denote tensile (shear) stress.

Although these are purely tensile strains, they can be transformed into shear strains by having the pasta structures oriented at an angle with respect to the simulation boundaries. 
Furthermore, nuclear pasta is not necessarily linearly elastic. Lasagna, for example, has transverse isotropy implying five independent elastic constants. Misalignment of the pasta structures with respect to the boundary and their time evolution makes it difficult to isolate these constants for all but the simplest cases. For most simulations, we observe a rotated stress tensor which mixes all $\sigma$ terms.

We strain at rates $\dot{\epsilon}=1\times10^{-7}$ and $2\times10^{-7}$ c/fm until a final strain $\epsilon_r$ along the axis of extension is achieved. This results in a compressional strain of $\epsilon_s=\epsilon_t=1/\sqrt[2]{2} \approx 0.29$ in the other two axes. These choices of strain rate does not affect our results. Although these deformations are much greater than those realized in astrophysical systems, they allow us to observe the breaking behavior of nuclear pasta.

\begin{figure*}
		\hrule \hrule \vspace{0.2cm}
	\subfloat[Tensile deformations pulling lasagna sheets apart.\label{fig1a}]{%
	\includegraphics[width=0.4\textwidth]{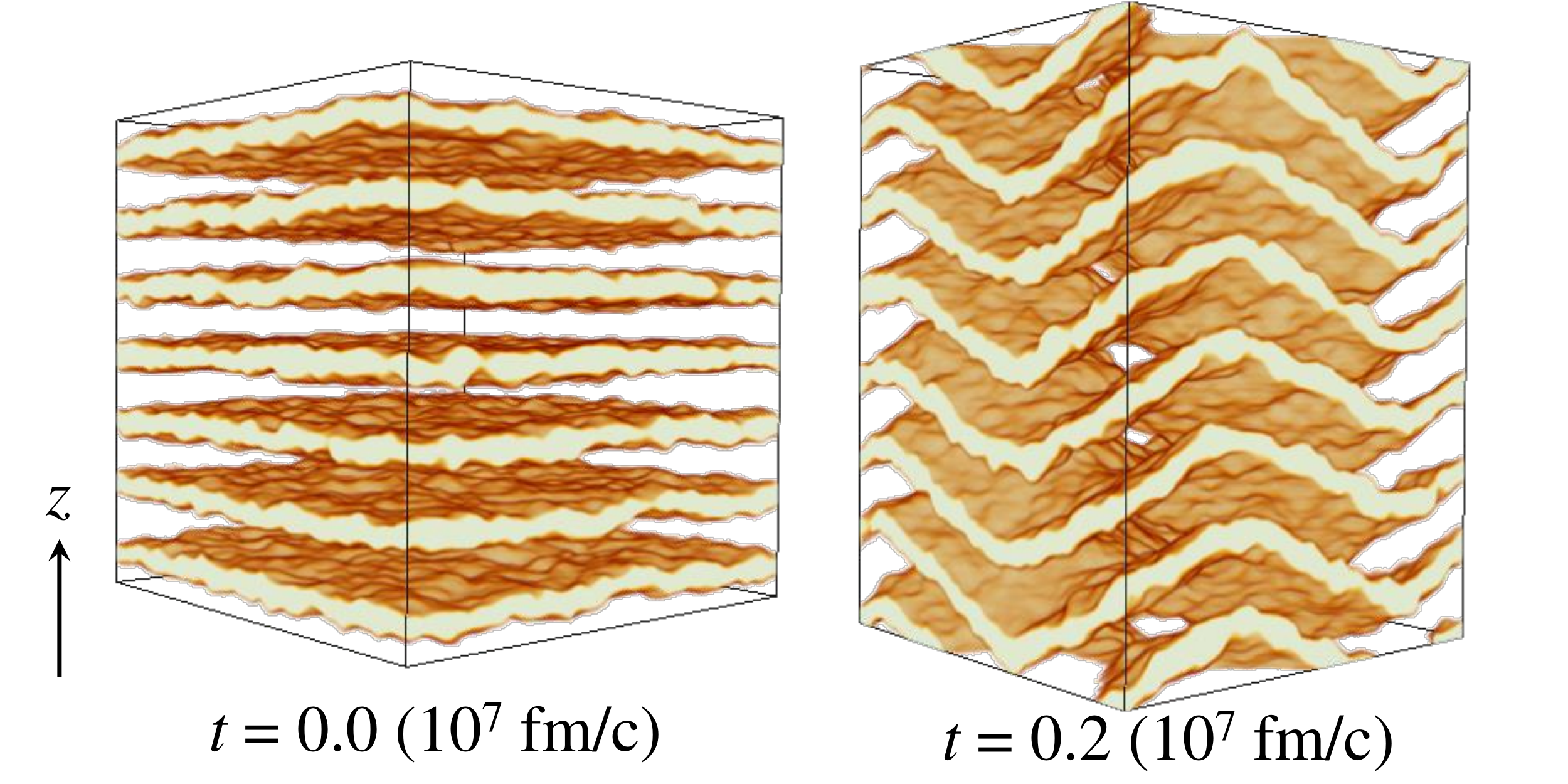}\qquad
  	\includegraphics[width=0.4\textwidth]{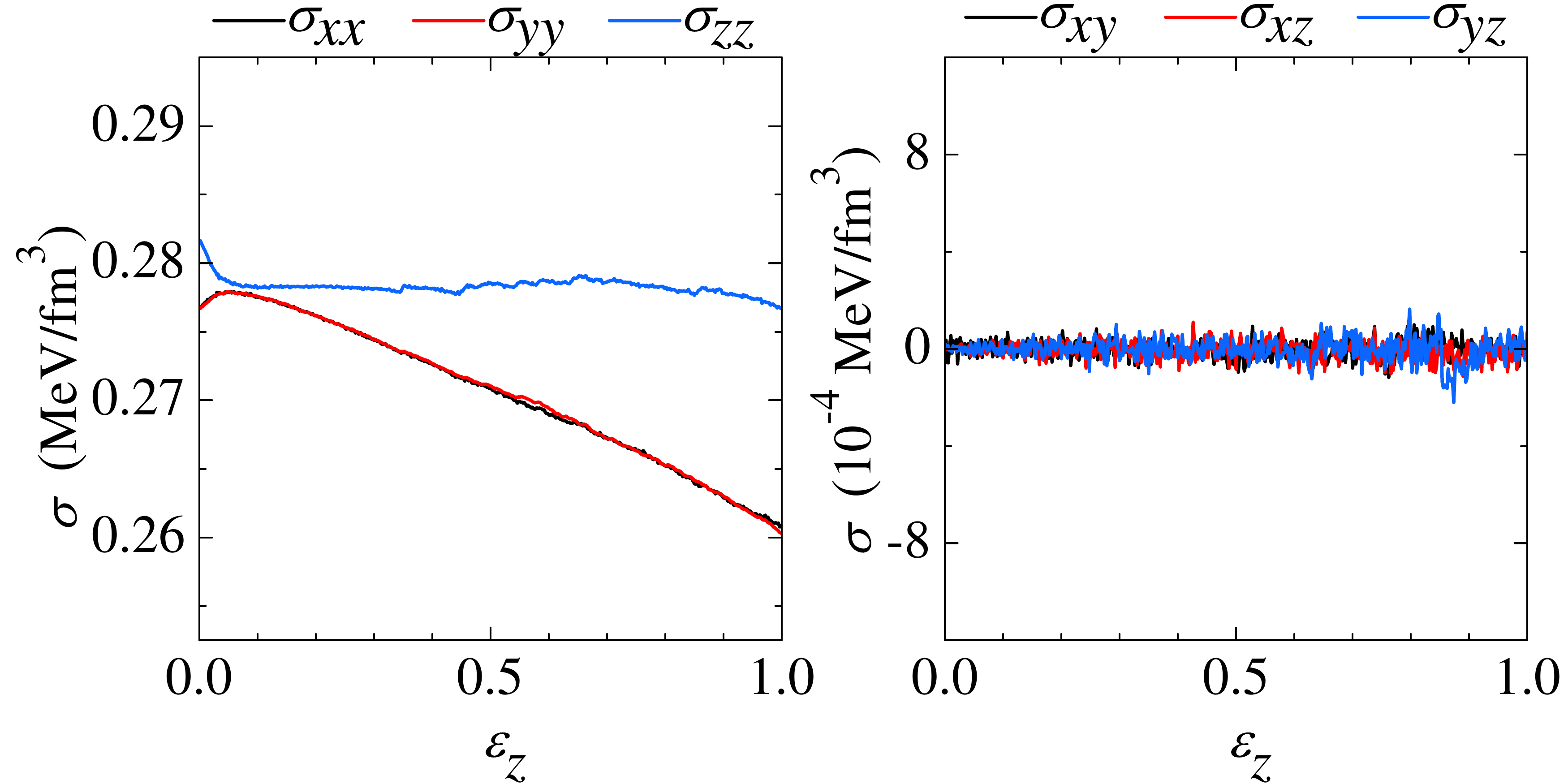}
	}
		\hrule \hrule \vspace{-0.2cm}
	\subfloat[Tensile deformation pulling lasagna sheets laterally while compressing them\label{fig1b}]{%
	\includegraphics[trim=0 10 0 10, clip, width=0.4\textwidth]{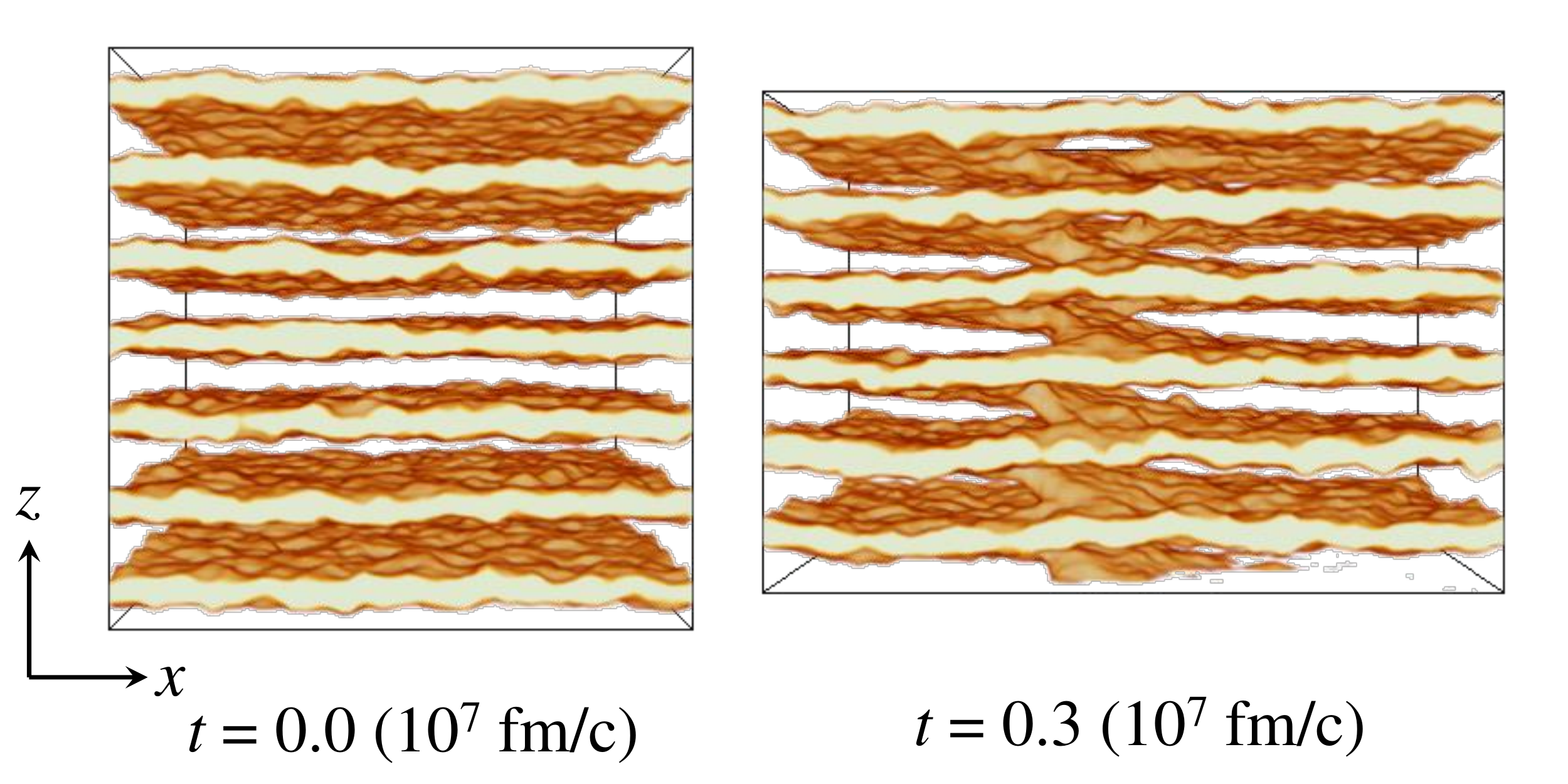}\qquad
  	\includegraphics[width=0.40\textwidth]{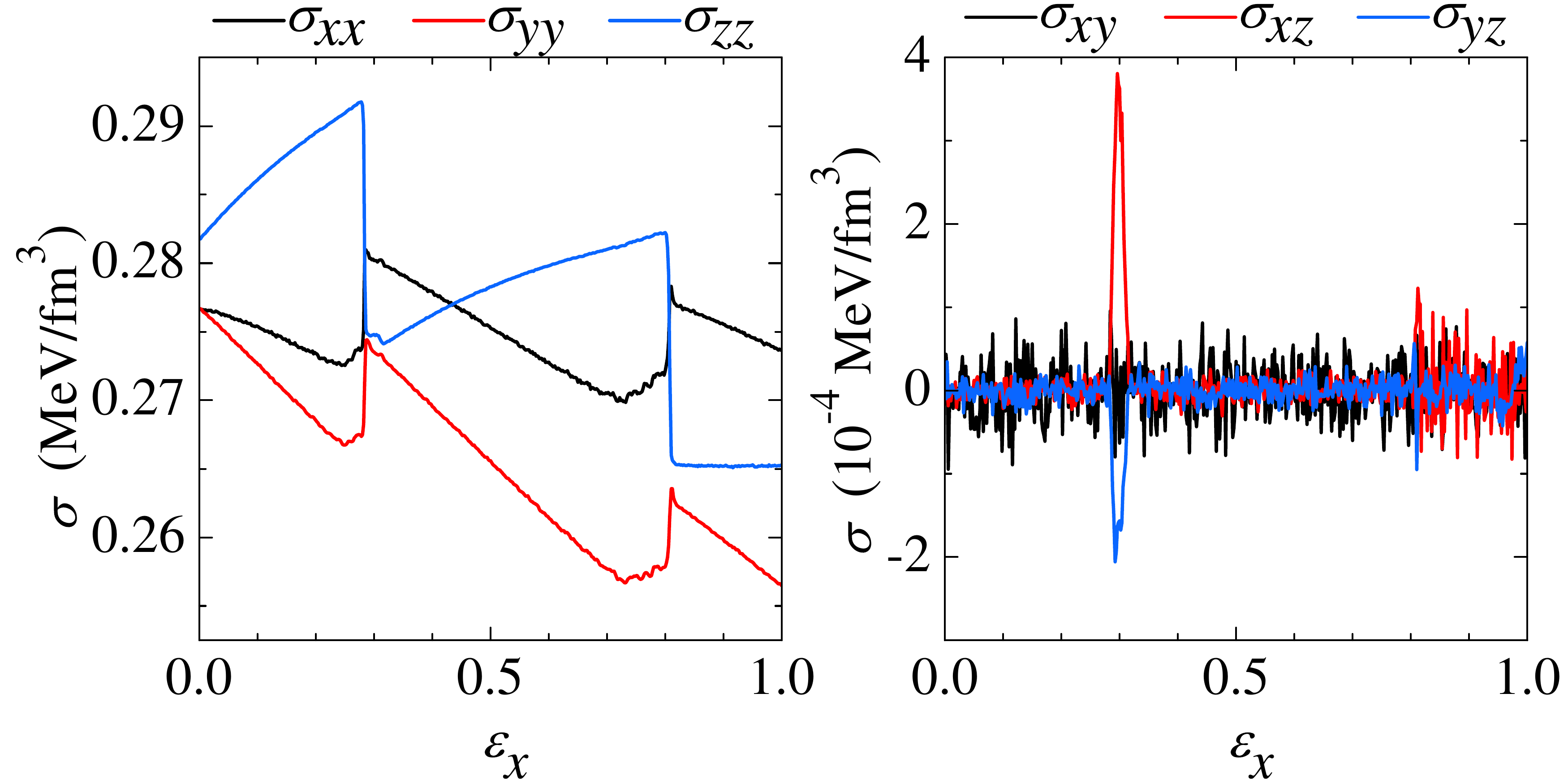}
	}
		\hrule \hrule \vspace{-0.2cm}
	\subfloat[Lasagna sheets experiencing both tensile and shear strains \label{fig1c}]{%
	\includegraphics[trim=0 10 0 0, clip, width=0.4\textwidth]{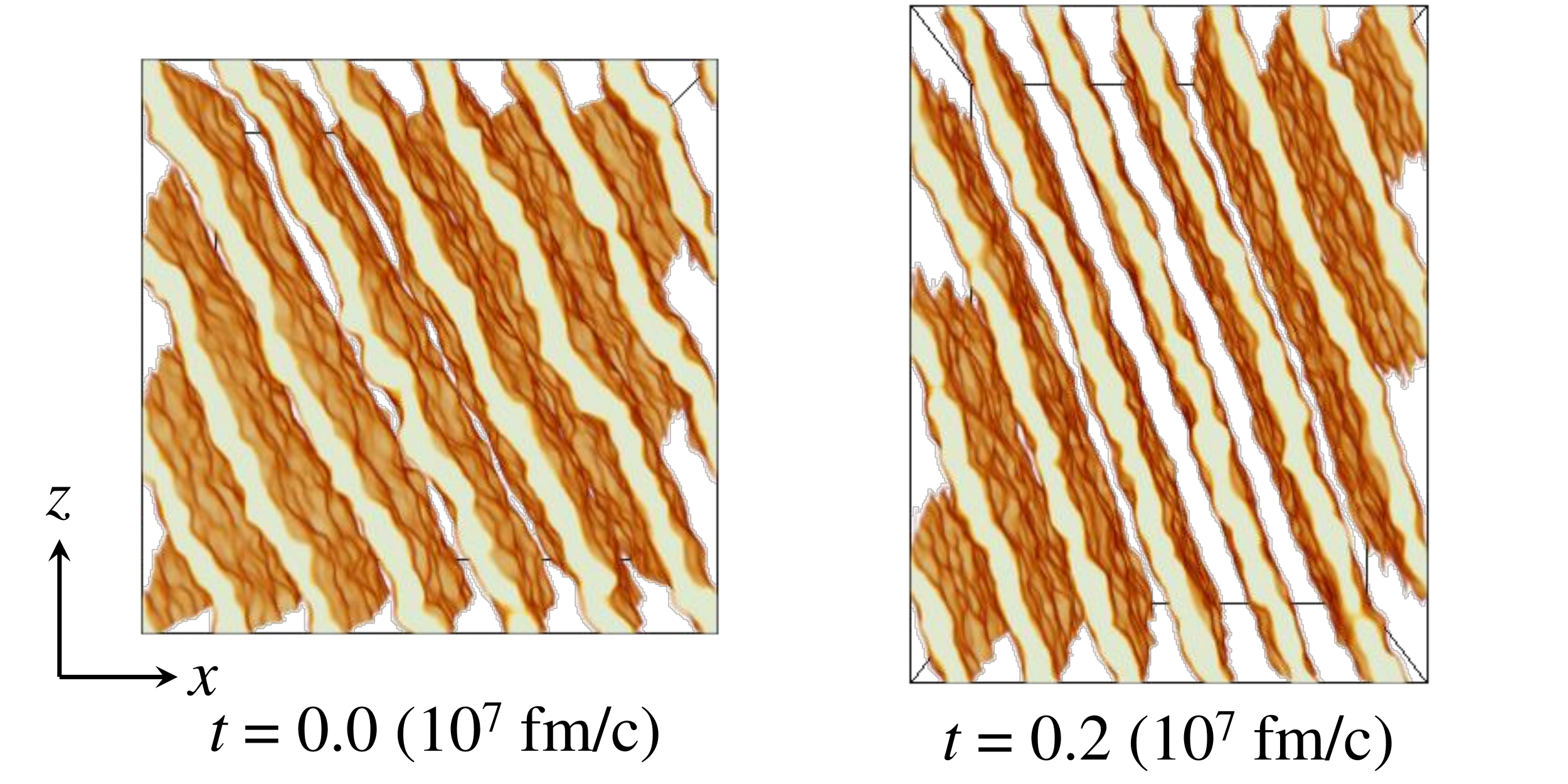}\qquad
  	\includegraphics[width=0.4\textwidth]{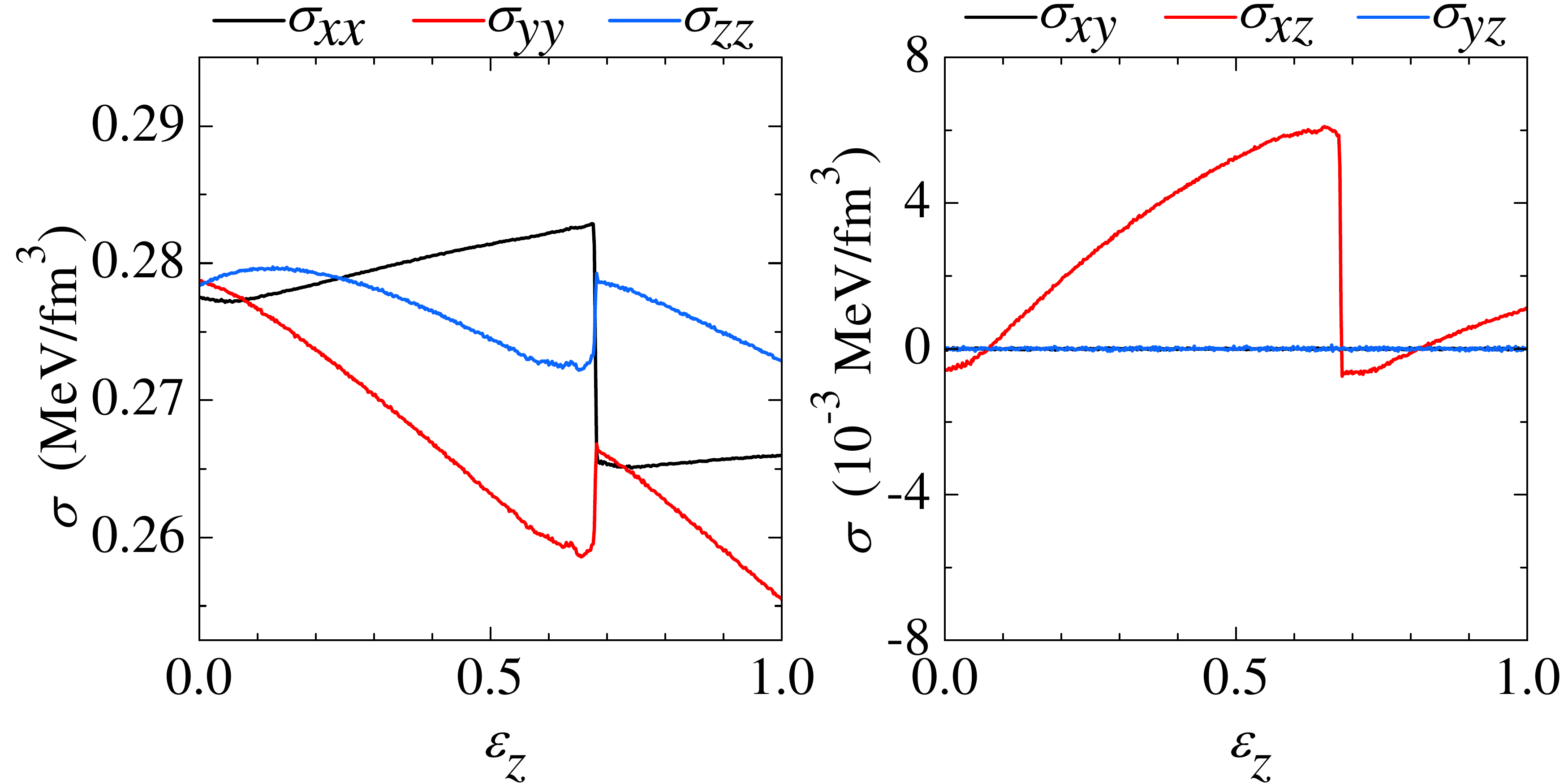}
	}
		\hrule \hrule \vspace{-0.2cm}
	\subfloat[Lasagna sheets with defects experiencing both tensile and shear strains, arrows show shear stress due to defects\label{fig1d}]{%
	\includegraphics[trim=0 10 0 10, clip, width=0.4\textwidth]{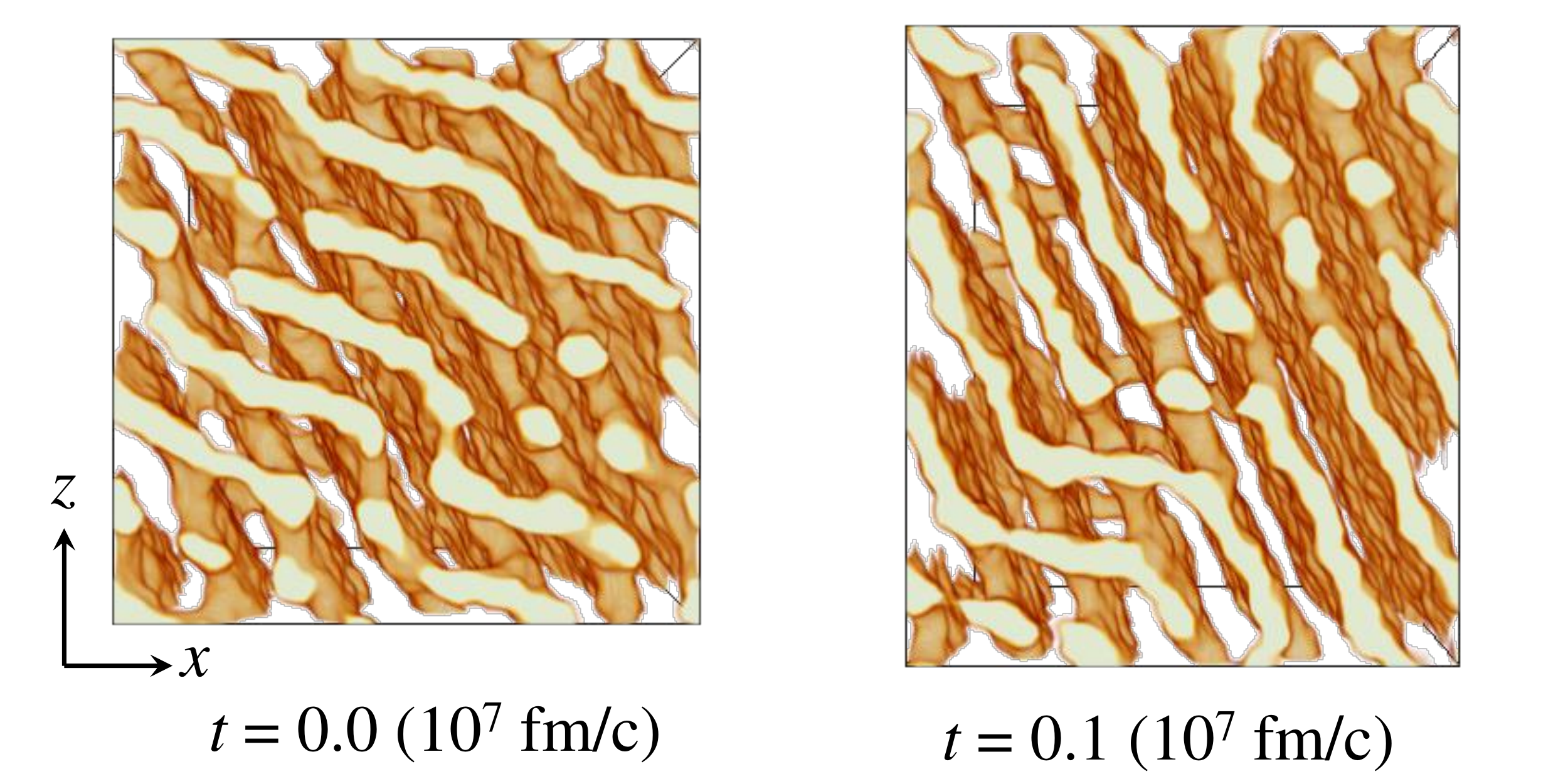}\qquad
  	\includegraphics[width=0.4\textwidth]{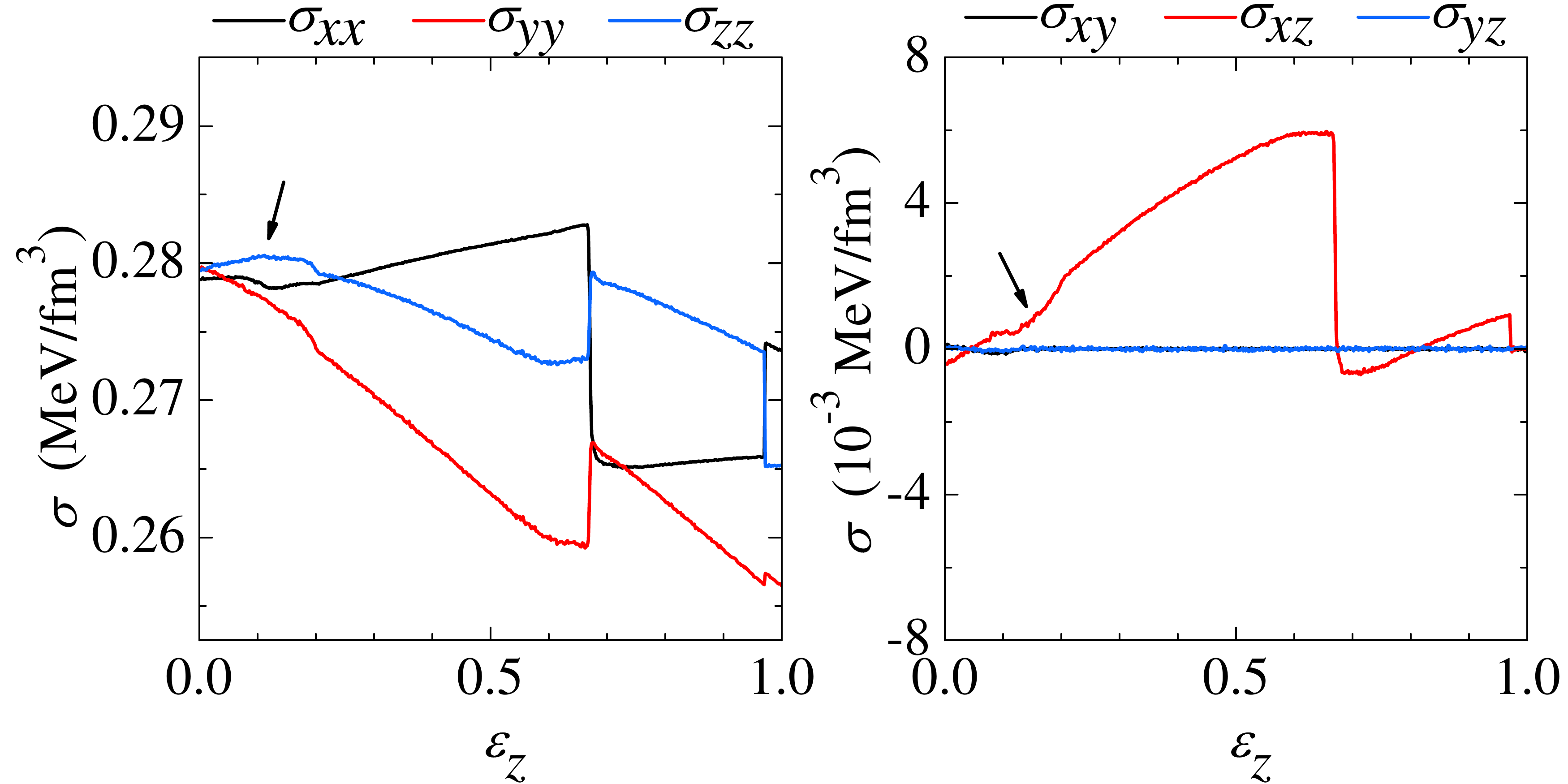}
	}
    \hrule \hrule
    \caption{Simulations deforming idealized nuclear pasta. Left to right, (1) initial configuration of the pasta, (2) deformed pasta at some later time, (3) tensile stresses vs.  extensional strain, and (4) shear stresses vs. extensional strain. Strain $\epsilon_r=l_r(t)-l_0$, see Eq. \eqref{strain}, is proportional to time. Animations avaiable in SM1-4 \cite{SM}. } 
\end{figure*}

First we deform `perfect' plates which are aligned with the $xy$ plane of the simulation volume \cite{PhysRevC.93.065806}. 
In the first case, Fig. \ref{fig1a}, we strain along the $z$ direction, pulling the plates apart, while contracting along the plate plane. For $\epsilon_z > 0.05$ the plates start to buckle to maintain their initial spacing. Tensile stress along $z$ remains constant from then on, while the stress in the plates decrease as the plates buckle. The plates deform continuously without breaking up till a strain $\epsilon_z=1$. As expected, no significant shear stresses are observed in this run.

In the second case, Fig. \ref{fig1b}, we strain along the $x$ axis, pulling along the plates and allowing their spacing to contract. Tensile stress along the $z$ direction grows as Coulomb pressure between plates increases. Meanwhile, in-plane stress decreases with the thinning of the plates. At $\epsilon_x \approx 0.3$ ($\epsilon_z \approx 0.13$), the plates break and the stresses abruptly change. The stresses reset and the process continues until another breaking event occurs at $\epsilon_x \approx 0.8$. 
To our knowledge, this breaking mechanism has not been described elsewhere in the literature and, thus, we describe it here in detail. 
We observe that (1) for small strains the spacing between plates decreases while the plates themselves become thiner. (2) Once the plates are thin enough, thermal fluctuations nucleate holes in them. (3) By displacing nucleons in the plate the holes relieves the Coulomb pressure between adjacent plates. (4) Adjacent plates become convex in the region of the holes. This partially fills the space vacated by the holes and allows the plates to increase their spacing locally, further reducing the system's pressure. (5) Convex lobes from the adjacent plates connect near the hole, forming short helicoidal bridges that connects adjacent plates. This allows connected plates to exchange nucleons. This process occurs in several locations simultaneously, allowing the plates to thicken and increase their interspacing locally. 
The bridges then drift through the plates until they are aligned, as seen in Fig. \ref{fig1b}, suggesting there is long range attraction between them as predicted by Guven \etal\,\cite{PhysRevLett.113.188101}. The filaments become thinner and disconnect as the plates reach their ideal thickness, resulting again in planar lasagna sheets but with one fewer plate spanning the simulation volume. 
During the first breaking event we observe that the shear stresses in the $xz$ and $yz$ planes are briefly nonzero, suggesting that filaments can support shear stresses between plates.

While the helicoidal bridges were short lived in this simulation, past work has identified stable configurations where filaments connecting plates are long lived \cite{PhysRevLett.114.031102}. These may support shear stresses between plates and, perhaps, even act as grain boundaries between domains where pasta plates have different orientations.

We also consider simulations of lasagna plates which are misaligned with the sides of the box allowing us to induce shear stresses in the material. In one simulation pasta plates are rotated in the $xz$ plane, Fig. \ref{fig1c}, while in another plates are similarly oriented but connected by stable and long lived filaments, Fig. \ref{fig1d}. Both systems have similar energy, suggesting that pasta has multiple available phases to it \cite{Newton2018}.

The system shown in Fig. \ref{fig1c} has three topologically disconnected plates due to the periodic boundary, while in the previous simulations all plates were topologically disconnected. These plates are free to adjust their inclination in response to the applied strain. They first break near $\epsilon_z \approx 0.68$, implying that pasta breaks less easily when the plates can adjust their inclination with respect to each other. The shear stress in the $xz$ plane grows approximately linearly until the break, transforming a tensile stress into a shear stress as expected. However, these shear stresses can be eliminated by applying a rotation matrix to the stress tensor whose angle is taken from the inclination of the plates, approximately recovering the stresses seen in Fig. \ref{fig1b}, as expected. This directly confirms that disconnected plates have near zero shear modulus between plates, as claimed by Pethick \& Potekhin \cite{PETHICK19987}.

In the run with inclined plates connected by helicoidal defects, Fig. \ref{fig1d}, the tensile deformation of the boundary produces a shear deformation of the plates due to their inclination. However, the filaments connecting the plates experience a shear stress as the plates slide, which is stronger in the $xz$ plane. The filaments break near $\epsilon_y \approx 0.2$, at which point the evolution proceeds identically to the case discussed in Fig. \ref{fig1c}.

The difference in $\sigma_{xz}$ at small strains seen between Fig. \ref{fig1c} and Fig. \ref{fig1d} cannot be eliminated by applying a rotation and is a clear indication of a shear modulus produced by the filaments. Their difference gives a maximum shear stress of $3 \times 10^{-4} \text{ MeV/fm}^3$; estimating the shear strain to be 0.1 gives us a shear modulus of $5 \times 10^{30} \text{ erg/cm}^3$. 
The rotation of the plates partially transforms the shear stress into a tensile stress at small strains, so this shear modulus may be a factor of two larger.

\begin{figure}[t]
\includegraphics[width=0.95\columnwidth]{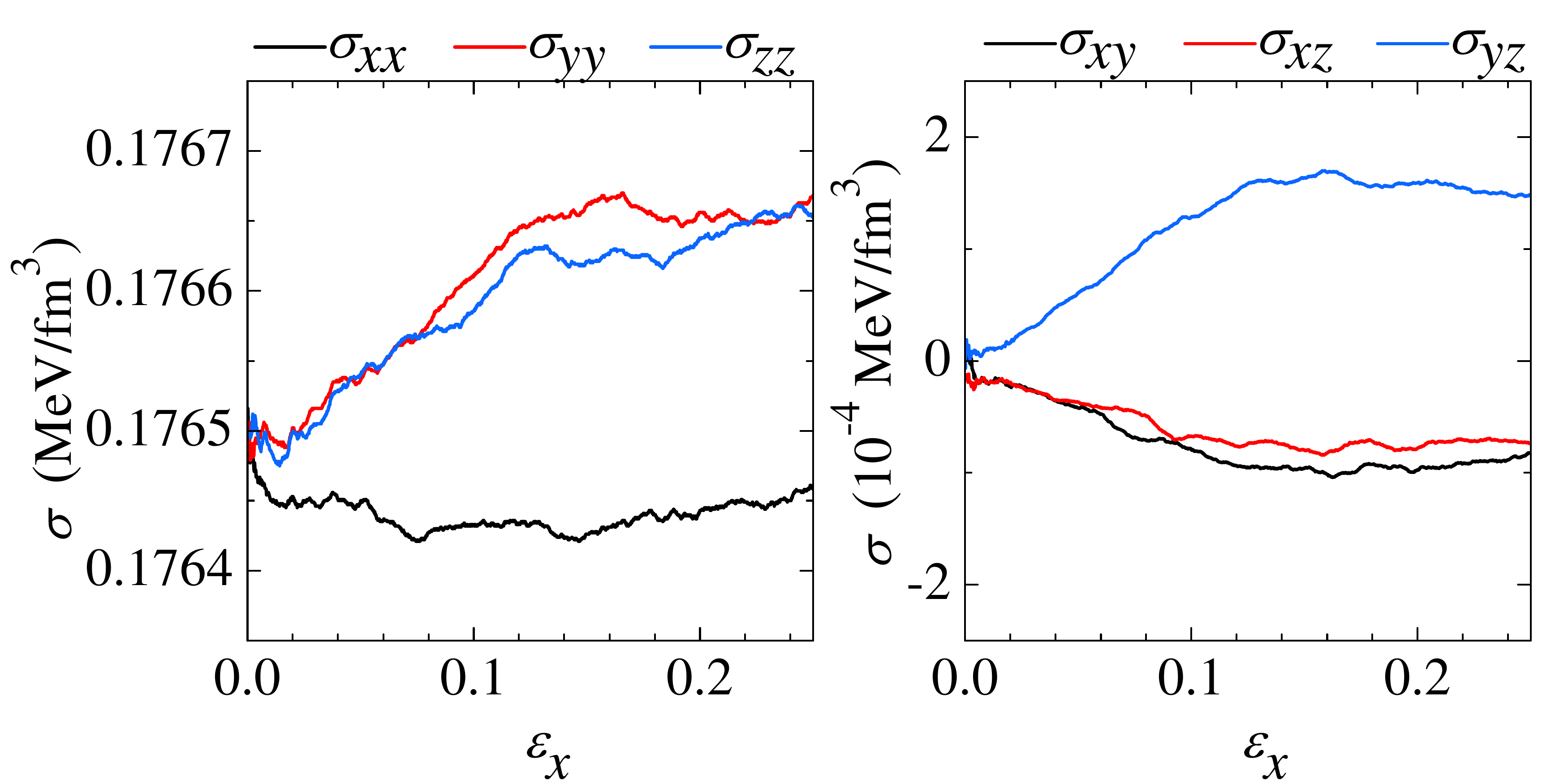}
\includegraphics[trim=0 5 15 0, clip, width=0.85\columnwidth]{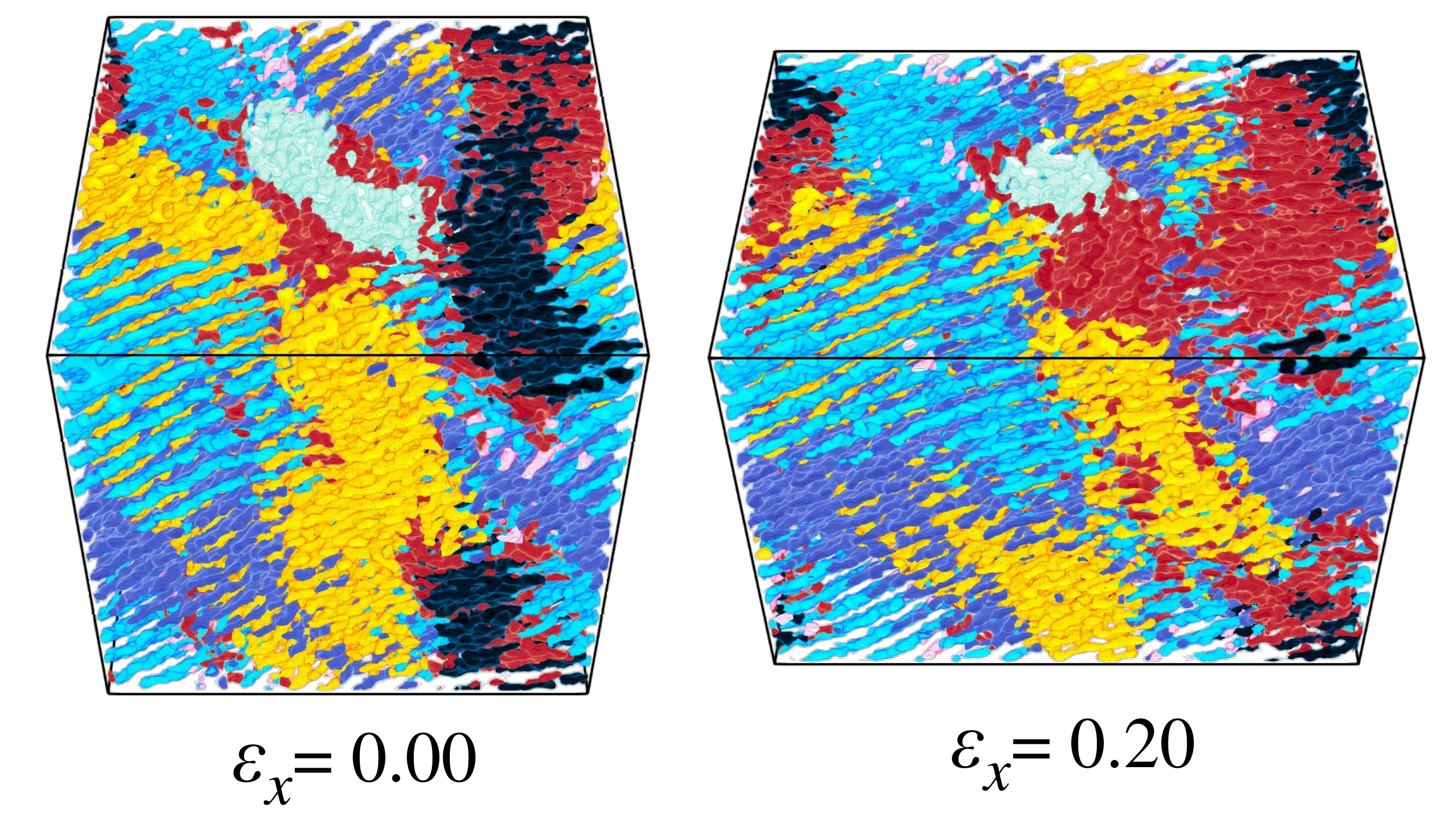}
\caption{Strain multidomain simulation of nuclear pasta. (Top) Tensile, left, and shear stresses, right. (Bottom) Evolution of domains, identified by color, from no strain ($\epsilon_x=0.0$), left, to $\epsilon_x=0.2$, right. Animations in SM5,6 \cite{SM}.  
}
\label{fig:4panel}
\end{figure}

With the elastic response of individual domains of nuclear pasta understood, we now consider a larger simulation with many domains. 
We present results of a simulation with 3\,276\,800 nucleons and proton fraction $Y_P = 0.3$. These parameters are known to produce parallel plates with a hexagonal arrangement of holes, the `waffle' phase \cite{PhysRevC.90.055805}. This proton fraction was chosen for two reasons. First, the simulation is less computationally expensive and allows us to evolve a larger number of particles. Second, the filamentary structure of the waffle phase allows us to study how its multiple domain structure affects pasta elasticity.

To achieve the initial condition for the strain run we evolved the system from a random initial condition for over 15,250,000 MD timesteps. This is the largest simulation of nuclear pasta yet and is discussed in detail in Ref. \cite{Schneider:18}. The system formed seven differently orientated domains, see Fig. \ref{fig:4panel}. The plates are misaligned at domain boundaries, but topologically connected by filaments. The relative volume fractions of domains remained approximately constant during this initial evolution, suggesting the domains are long lived. 

This simulation was strained at a rate $\dot{\epsilon_x} =2 \times 10^{-7}$ c/fm until $\epsilon_x = 0.25$. The stresses are shown in Fig. \ref{fig:4panel}. No breaking was observed. Shear and tensile stresses with components perpendicular to the strain direction $x$ grow approximately linearly until $\epsilon_x \sim 0.12$ and then flatten out. Meanwhile, stresses with $x$ components decrease slightly before flattening out. This is expected for a linear isotropic material which is misaligned with the strain axis. Another simulation was strained at $\dot{\epsilon}_x = 1 \times 10^{-6}$ c/fm until $\epsilon_x = 0.77$ and finds similar results. 
While it is not possible to extract individual elastic constants from this simulation, the stresses grow comparably to our $Y_p=40\%$ simulations, suggesting they may be similar.

In Fig. \ref{fig:4panel} we show our simulation for two different strains, $\epsilon_x = 0.0$ and $0.2$. 
At their boundaries, domains are connected by filaments similar to the helicoidal defects that connect lasagna plates and considered above. When strained, domains move relative to each other and slide with only limited resistance. At the boundaries nucleons flow freely between adjacent domains, rearranging nuclear matter and allowing domains to shrink or grow in response to the applied stress. This prevents the large stresses that result in catastrophic failures from building up. As nuclear pasta in the neutron star crust is expected to form domains, the inner crust may not break for realistic astrophysical strains.

\textit{Discussion.} We find, using MD simulations, that the analytic model of Pethick \& Potekhin describes well the qualitative elastic response of idealized lasagna plates to large deformations. Idealized nuclear pasta is strong in our model, possibly with a shear modulus as large as $10^{31} \,\text{erg/cm}^{3}$. This is comparable to the strength of the outer crust extrapolated to the high densities at the crust-core boundary, which may make nuclear pasta the strongest material in the known universe. 

For comparison, a typical nucleus predicted at pasta densities by Haensel \& Zdunik is $^{88}$Ti (Z=22) with free neutron fraction $X_n=0.80$ \cite{Haensel1990}. Using the effective shear modulus of Horowitz \& Hughto $\mu_{eff} \sim 0.11 n Z^2 e^2 / a$ (ion number density $n$, $a=[3/(4 \pi n)]^{1/3}$) we find $\mu = 1.1 \times 10^{30}$ \cite{Horowitz2008}. We emphasize that the large shear modulus and breaking strain for our pasta simulations are model dependent. Our simulations use proton fractions much larger than those expected in the inner crust of NSs. A lower proton fraction likely reduces the shear modulus, perhaps to $10^{30} \text{erg/cm}^{3}$. This motivates further work on pasta elasticity, especially the role that superfluidity may play, which is beyond the scope of our simulations \cite{2018arXiv180306254K}.

We explicitly describe the breaking mechanism, and find that the breaking strain of idealized nuclear pasta in the lasagna and waffle phases is large in our model, perhaps $\epsilon \approx 0.3$. This suggests that the ion crust breaks significantly earlier than the pasta. 
This may have consequences for crust breaking in a variety of systems, such as resonant shattering flares during neutron star mergers or magnetar outbursts. Additionally, the large strength and density of nuclear pasta predicted by this work suggests that neutron stars may support large `buried' mountains in the inner crust. These could be an efficient source of continuous gravitational waves, and further motivates searches by LIGO, Virgo, and other GW detectors.

While we can describe the elastic response of some simple astromaterials in this work, the large variety of shapes that nuclear pasta adopts and the variety of elastic responses to induced strains observed here suggests that a theory of pasta elasticity may be difficult to formulate analytically, and motivates future work exploring pasta elasticity with MD, such as isolating the elastic constants of the various phases of pasta.

\textit{Acknowledgements.} M.C. is a CITA National Fellow. A.S.S. is supported by the National Science 
Foundation under award No. AST-1333520 and CAREER PHY-1151197. C.J.H. supported in part by US Department of Energy grants DE-FG02-87ER40365 (Indiana University) and DE - SC0018083 (NUCLEI SciDAC-4 Collaboration). This research was supported in part by Lilly Endowment, Inc., through its support for the Indiana University Pervasive Technology Institute.
We thank A. Cumming and A. Chugnuov for conversation.

\end{document}